\documentstyle[12pt,aaspp4]{article}

\lefthead{}
\righthead{LMC X-1}

\begin{document}

\title{$RXTE$ OBSERVATIONS OF LMC X-1}

\author{P.C. Schmidtke, A.L. Ponder, and A.P. Cowley}
\affil{Department of Physics \& Astronomy, Arizona State University,
Tempe, AZ, 85287-1504} 

\begin{abstract}

The luminous X-ray binary and black-hole candidate LMC X-1 has been
observed with the Rossi X-ray Timing Explorer ($RXTE$) to search for
quasi-periodic oscillations (QPO), previously reported in its high state.
The source was observed monthly in a series of nine observations. 
Analysis of the temporal variations shows no evidence for QPO or other
periodic changes, but correlations between the high-energy light curve and
hardness of the spectrum are described.  Spectral fits with two-component 
models demonstrate that the hardness variations come from changes in the 
intensity of the high-energy power-law tail.

\end{abstract}


\section{INTRODUCTION}
    
LMC X-1 is one of four extremely luminous ($>10^{38}$ ergs s$^{-1}$) X-ray
binaries in the Large Magellanic Cloud.  The source has long been known to
show a rather soft X-ray spectrum ($kT\sim2.7$ keV; Markert \& Clark 1975)
and irregular X-ray variability by at least a factor of three (Griffiths
\& Seward 1977, Johnston, Bradt, \& Doxsey 1979).  However, its optical
identification was uncertain for many years.  The source lies within the bright
emission nebula N159 (Henize 1956) containing many early-type stars (see
finding chart of region in Cowley, Crampton, \& Hutchings 1978).  Based on the
rather uncertain X-ray position, it was originally thought that the B5
supergiant R148 ($V\sim12.5$) was the optical counterpart, although the
peculiar O7 III star (Star \#32 of Cowley et al.) $6^{\prime\prime}$ away could
not be ruled out.  Recent analysis of $ROSAT$ High-Resolution Imager data
has confirmed that Star \#32 ($V\sim14.8$) is the most probable identification
(Cowley et al.\ 1995).  Spectroscopic studies of this star reveal an orbital
period near 4 days and a probable mass for the compact star of
$M_X\sim4M_{\odot}$ (Hutchings, Crampton, \& Cowley 1983, Hutchings et
al.\ 1987), making it a strong black-hole candidate. 

Because LMC X-1 is so luminous, its X-ray properties were studied even with
the earliest X-ray detectors.  White \& Marshall (1984) discussed the
complex character of its X-ray spectrum, showing that it could not be
described by a simple model because of the high-energy excess above 3 keV.
Thus a two-component model with a soft thermal spectrum and plus a hard
high-energy tail was needed.  They pointed out the similarity in spectral
properties of LMC X-1 to another black-hole candidate, LMC X-3, and they
were the first to recognize that an unusually soft X-ray spectrum may be a
reliable signature of a black-hole candidate. 

Using {\it Ginga} data, Ebisawa, Mitsuda, \& Inoue (1989) also found a
two-component model was needed to fit the spectrum: an ultrasoft blackbody
($kT\sim0.8$ keV) and a hard power law (photon index $\sim$2.5).  Later,
Schlegel et al.\ (1994a, 1994b) modeled the spectrum, based on Broad Band
X-ray Telescope (BBXRT) data, finding results similar to those of Ebisawa
et al. 

Ebisawa et al.\ undertook a timing analysis of their {\it Ginga} data and
found quasi-periodic oscillations (QPO) with a peak frequency of 0.0751 Hz
in one of the observations.  They concluded that the QPO came from the
hard-tail component, which was unusually bright at the time in comparison
to the thermal component.  QPO were not found in the BBXRT data (Schlegel
et al.), confirming the transient nature of these aperiodic signals.  Since
the presence of QPO might be related to the spectral state of the system,
we undertook a series of observations with Rossi X-ray Timing Explorer ($RXTE$)
over a period of nine months to search for QPO and to further define their
origin.

\section{OBSERVATIONS AND DATA ANALYSIS} 

A series of nine observations were made between 1996 February and 1996
October.  Table 1 lists the details of the observational data, in reverse
time order, following the numbering system of the $RXTE$ Science Data
Center.  Only data from the Proportional Counter Array (PCA) were analyzed
due to the very low net count rate for events with energies $>$16 keV.
Each observation was broken into two or three segments due to
Earth occultations or passages through the South Atlantic Anomaly.  In
1996 March, two proportional counter units (PCU) were found to be
discharging, making it necessary to shut them down periodically.  The
result was that some observations were made with only three of the five
PCU operating (observations \#4, \#6, and \#7), as noted in Table 1.
Observation \#3 experienced a shutdown of two PCU during the second half
of the exposure, so it was necessary to analyze this observation in two
parts (\#3A and \#3B).

\subsection{Light Curves and Hardness Ratios}

Background-subtracted light curves, using 16-s time bins, were derived for
the 2--5.9 keV and 5.9--15.9 keV energy ranges.  The spectral hardness within
each bin was found by calculating the ratio of counts, 
$HR =$ (5.9--15.9 keV)/(2--5.9 keV).
Typical values for the hardness ratio are $\sim$0.1--0.2; mean values for each
observation are listed in Table 1.  The light and $HR$ curves for observations
\#2, \#3B and \#4 are shown in Fig.\ 1.
The selected observations span a wide range of source intensity and hardness.
In observation \#3B (`average' count rate; lowest 
$HR$) LMC X-1 was in a quiescent mode with minimal flickering, while in both
\#2 (second highest count rate; highest $HR$) and \#4 (lowest count rate;
`average' $HR$) the source varied rapidly on a time scale of minutes.
Examination of all of the observations reveals no dependence of the source's
flickering characteristics on either the count rate or hardness.  However, for
each observation the plot of hardness ratio is similar to that of the
high-energy count rate, implying the ratio is driven primarily by the number of
5.9--15.9 keV counts.  $Ginga$ data for the black hole
candidate GX{\thinspace}339$-$4 in its very high state (Miyamoto et al.\ 1991)
show the same behavior (see their Fig.\ 2).  While in this very high state,
the source is 2--3$\times$ more intense and shows significantly enhanced
variability (on timescales of minutes) compared to when it is in a (relatively
quiescient) high state (Makishima et al.\ 1986).  Although the $RXTE$
observations also show various degrees of short-term variability, there is no
clear indication that LMC X-1 enters a very high spectral state.

To test the dependence of $HR$ on the source count rate, we used data from all
observations to construct the hardness-intensity diagrams shown in Fig.\ 2.
The plot of $HR$ versus low-energy count rate has no significant relationship
between the parameters.  However, the plot of $HR$ versus 5.9--15.9 keV count
rate exhibits a very strong correlation which is also present in each of the
individual observations.  This dependence confirms the trend inferred from
examination of the simple light and hardness-ratio curves.

We examined the data for time lags between high and low-energy photons by
calculating the cross-correlation of the two light curves.  A small positive
lag, in the sense that high-energy photons tend to arrive after their
low-energy counterparts, was found for all observations.  However, the mean lag
time in each case was comparable to the 0.025-s step used in the calculations,
so that the results are not conclusive.

\subsection{Power Density Spectra and Quasi-periodic Oscillations}

To further investigate the temporal variations, power density spectra (PDS)
were constructed using the background-subtracted data, rebinned at 0.125-s
resolution.  Within an observation, the Fourier transformation was calculated
for each 512-s data segment (4096 points), covering the frequency range
0.002--4 Hz.  Typically, each PDS represents the average of $\sim$11 individual
transforms.  Two sets of PDS were made, one covering the entire PCA energy
range (2--60 keV) and the other restricted to primarily those photons in the
hard spectral tail (5.9--15.9 keV).  No significant differences between these
were found.  In Fig.\ 3 we present the power density spectrum (5.9--15.9
keV) for observation \#7.  The power density increases at low frequency (i.e.,
red noise) and can be modeled in terms of a power law plus a constant (i.e.,
power $= \nu^{-\alpha} + C$, where $\alpha$ is the power index).  Fitted values
of the power index, for both the 5.9--15.9 keV and 2--60 keV PDS, are listed in
Table 2. The average indices are $\alpha=1.06$ and 0.87 for the restricted and
full energy ranges, respectively.  There is no significant dependence of the
modeled power index with either the hardness ratio or the source count
rate.  For comparison, Ebisawa et al.\ (1989) measured an index of 0.81 in a
$Ginga$ observation taken when LMC X-1 was in an X-ray bright state.

If present, quasi-periodic oscillations would appear as a broad peak in
the PDS.  None of our observations shows this signature, although Ebisawa et
al.\ found strong QPO near 0.0751 Hz in some of the $Ginga$ data, with weaker
QPO around 0.142 Hz.  (Most likely, this other peak is the second harmonic.)
Upper limits for the amplitude of QPO signals in the $RXTE$ observations were
found by calculating the percent r.m.s.\ variation of the average excess power
(after subtracting the fitted model) between 0.05 and 0.10 Hz.  Typical
3$\sigma$ limits are $\sim$0.8\% for the 2--60 keV PDS, as listed in Table 2.
For comparison, the amplitudes of the primary and secondary $Ginga$ QPO peaks
were 2.9\% r.m.s.\ and 1.8\% r.m.s., respectively.  Hence, the power found in
the $Ginga$ QPO was much larger than the upper limits placed on the $RXTE$
data.  Clearly, the appearance of QPO in LMC X-1 is a transient phenomenon,
as further documented by the lack measurable QPO in BBXRT observations of
this source (Schlegel et al.\ 1994a, 1994b).

\subsection{Spectral Fits}

The X-ray energy spectrum of LMC X-1 was separately fit by two multi-component
models over the energy range 3.6--15.9 keV.  The soft thermal component was
represented by both a blackbody and by a multicolor disk model (Mitsuda
et al.\ 1984; Makishima et al.\ 1986), while the high-energy excess was always
accounted for using a simple power law.  Within limitations imposed by an
uncertain background subtraction at low energies in the $RXTE$ data, both
combinations yield acceptable fits.  The intervening hydrogen absorption was
fixed at $N_H=0.6{\times}10^{22}$ cm$^{-2}$ in all cases, since the
derivation of $N_H$ as a free parameter is very sensitive to errors in
background subtraction.  Mean values for the fitted parameters, which are
similar to those of Ebisawa et al.\ (1989) and Schlegel et al.\ (1994a),
are summarized in Table 3.  Since the hardness ratio
depends on source intensity, further modeling was performed on data selected
by PCA count rate count (i.e., 130--140 cnts s$^{-1}$, 140--150 cnts s$^{-1}$,
etc.).  The results of this fitting for the blackbody plus power-law model are
shown in Fig.\ 4.  Within each observation, we find a trend for an increasing
`scale' of the power law (i.e., comparable to intensity provided the photon
index is constant) with increasing source count rate (3.6--15.9 keV).  The
remaining parameters are nearly constant within a given observation, although
a slight increase in temperature of the thermal component may be present at
higher count rates.  We note that the fitted parameters are highly coupled.
If the temperature of the soft component were held constant, then the
dependence of the power-law scale on count rate would become even more
pronounced.

\subsection{Long-term and Phase-related Variations}

The All Sky Monitor (ASM) aboard $RXTE$ provides nearly continuous monitoring
of the entire sky, producing long-term light curves for bright, persistent
X-ray sources as well as discovering new transient sources.  We have examined
the LMC X-1 data in this public archive.  They compare well with the mean
count rates for our pointed observations, as presented in this paper.  Overall,
the source varied by about a factor of two through the nine months during
which our observations were taken.  Analysis of $\sim$3500 ASM data points
taken over 76 weeks of observing does not reveal any significant periods in
the range 0.5--20 days, including at the optically determined orbital period
of 4.22 days.  This result has been confirmed by scientists on the instrument
team (Levine 1997). 

In addition to searching for periods in the ASM observations, we have
folded our $RXTE$ data on the 4.22-day orbital period.  The phase coverage
is poor because the observations were only obtained at nine discrete
times.  As it turned out, no observations were obtained between
spectroscopic phases 0.9 to 1.3, leaving a large gap in the phase plots.
From the folded data there is no clear indication of any trends related to
the orbital period either in the count rate or the hardness ratio.  Additional
data obtained at more closely spaced intervals are need to determine if
any small orbital modulation or other signature is present in the X-ray 
data.

\section{SUMMARY}

We obtained a series of $RXTE$ observations of the black-hole candidate
LMC X-1 to search for quasi-periodic oscillations and examine their
behavior with source intensity.  No QPO were detected, although Ebisawa et
al.\ had detected them in LMC X-1 when the source was in a bright state.
Comparison of the source intensity during the $RXTE$ and {\it Ginga}
observations suggests that LMC X-1 never became as bright during our
$RXTE$ observations as when QPO were found in {\it Ginga} data.  However,
because of the uncertainty of the background subtraction with the $RXTE$
data, it is difficult to compare these data sets.  In addition, no other
periodic behavior was found, either on the orbital period or any other
in the range of 0.5-20 days. 

Fits of the X-ray spectral energy distribution require a two-component
model, and the values derived are similar to those found by previous
workers.  A strong correlation was detected between the hardness ratio and
the high-energy count rate, indicating that changes in the X-ray spectrum
of LMC X-1 come from the high-energy tail.

\acknowledgments
We wish to thank the staff at the $RXTE$ Guest Observer Facility.  We 
also acknowledge support from NASA. 

\clearpage

\begin{table}
\caption[]{Summary of {\it RXTE} Observations}
\begin{flushleft}
\begin{tabular}{lccccc}
\hline
\hline
Obs. & ~~Date & Start Time & Exp.\ Time & Count Rate$^a$ & Hardness \\
~\# & yy/mm/dd & UT & s & cnts s$^{-1}$ & Ratio$^b$ \\
\hline
~1  & 96/10/04 & 12:35:12 & 4912 & 155.6$\pm$0.3     & 0.15$\pm0.03$ \\
~2  & 96/09/06 & 21:45:52 & 6880 & 162.2$\pm$0.3     & 0.23$\pm0.05$ \\
~3B & 96/08/01 & 22:12:32 & 2880 & 151.0$\pm$0.3$^c$ & 0.10$\pm0.02$ \\
~3A & 96/08/01 & 21:02:56 & 3664 & 160.6$\pm$0.3     & 0.13$\pm0.02$ \\
~4  & 96/07/05 & 01:59:44 & 5328 & 135.2$\pm$0.4$^c$ & 0.15$\pm0.04$ \\
~5  & 96/06/09 & 20:22:24 & 6800 & 148.8$\pm$0.3     & 0.14$\pm0.03$ \\
~6  & 96/05/18 & 07:32:48 & 6656 & 159.2$\pm$0.4$^c$ & 0.15$\pm0.03$ \\
~7  & 96/04/18 & 06:45:52 & 5840 & 168.2$\pm$0.4$^c$ & 0.15$\pm0.03$ \\
~8  & 96/03/08 & 05:39:12 & 4976 & 148.1$\pm$0.3     & 0.22$\pm0.06$ \\
~9  & 96/02/10 & 05:31:12 & 9504 & 155.1$\pm$0.3     & 0.18$\pm0.03$ \\
\hline

\end{tabular}
\end{flushleft}
$^a$Background-subtracted PCA count rate (2--60 keV) \\
$^b$Hardness ratio $=$ (5.9--15.9 keV count rate)/(2--5.9 keV count rate) \\
$^c$Corrected to account for shutdown of two PCU \\
\end{table}

\begin{table}
\caption[]{Power Density Spectrum Parameters}
\begin{flushleft}
\begin{tabular}{lccc}
\hline
\hline
Obs. & Power Index$^a$ & Power Index$^a$ & QPO Power$^b$ \\
~\# & 5.9--15.9 keV & 2--60 keV & 2--60 keV \\
\hline
~1  & $0.83\pm0.07$ & $0.63\pm0.05$ & $<0.4\%$ \\
~2  & $0.99\pm0.02$ & $0.88\pm0.01$ & $<0.2\%$ \\
~3B & $1.09\pm0.09$ & $1.01\pm0.04$ & $<0.8\%$ \\
~3A & $1.17\pm0.12$ & $1.17\pm0.06$ & $<1.0\%$ \\
~4  & $0.78\pm0.03$ & $0.64\pm0.01$ & $<1.8\%$ \\
~5  & $0.94\pm0.09$ & $0.93\pm0.05$ & $<0.5\%$ \\
~6  & $1.44\pm0.07$ & $1.09\pm0.03$ & $<0.6\%$ \\
~7  & $0.98\pm0.03$ & $1.02\pm0.01$ & $<1.6\%$ \\
~8  & $1.49\pm0.03$ & $0.85\pm0.01$ & $<0.9\%$ \\
~9  & $0.86\pm0.04$ & $0.52\pm0.03$ & $<0.2\%$ \\
\hline

\end{tabular}
\end{flushleft}
$^a$Modeled using $\nu^{-\alpha} + C$, where $\alpha$ is the power index \\
$^b$3$\sigma$ upper limit for r.m.s.\ variation of average excess power
between 0.05--0.10 Hz \\
\end{table}

\begin{table}
\caption[]{Mean Spectral Fitting Parameters}
\begin{flushleft}
\begin{tabular}{ll}
\hline
\hline

\underbar{Model: Blackbody plus power law} \\
~~~~ Radius of blackbody$^a$ & 40 km \\
~~~~ Temperature of blackbody & 0.88 keV \\
~~~~ Photon index of power law & 2.44 \\
~~~~ Scale of power law & $2.08{\times}10^{-4}$ photons keV$^{-1}$ cm$^{-2}$ s$^{-1}$ at 10 keV \\
~~~~ $N_H$ & $0.6{\times}10^{22}$ atoms cm$^{-2}$ (fixed) \\

\underbar{Model: Multicolor disk plus power law} \\
~~~~ Projected size of inner disk radius$^{a,b}$ & 22 km \\
~~~~ Temperature of inner disk radius & 1.17 keV \\
~~~~ Photon index of power law & 1.45 \\
~~~~ Scale of power law & $1.42{\times}10^{-4}$ photons keV$^{-1}$ cm$^{-2}$ s$^{-1}$ at 10 keV \\
~~~~ $N_H$ & $0.6{\times}10^{22}$ atoms cm$^{-2}$ (fixed) \\

\hline

\end{tabular}
\end{flushleft}
$^a$For an assumed distance of 50 kpc \\
$^b$R$_{in}$cos$^{\rm{1/2}}\theta$ \\
\end{table}

\clearpage

\begin{figure}
\caption{$RXTE$ data for observations \#2, \#3B, and \#4 of LMC X-1; the top
and middle panels are X-ray light curves for the energy ranges 2--5.9 keV and
5.9--15.9 keV, respectively.  The bottom panels show the hardness ratio
(5.9--15.9 keV/2--5.9 keV).  Note the similarity of the high-energy light
curves and the hardness ratio.  The other observations have similar
characteristics.  The short-term variability does not appear to depend on
either source intensity or hardness.}
\end{figure}

\begin{figure}
\caption{Hardness ratio versus count rate in the 2--5.9 keV and 5.9--15.9 keV
energy ranges for all $RXTE$ observations of LMC X-1.  Although there is no
relationship between hardness and the low-energy counts, the ratio is strongly
correlated with the high-energy count rate.  This correlation is present in
each observation.}
\end{figure}

\begin{figure}
\caption{Power density spectrum of LMC X-1 for $RXTE$ observation \#7 using
0.125-s bins over the energy range 5.9--15.9 keV.  The best-fit curve is a
power law plus a constant (i.e., power $= \nu^{-\alpha} + C$).  There is no
evidence for QPO in this or any of the other PDS within the frequency range
0.002--4 Hz.} 
\end{figure}

\begin{figure}
\caption{Parameters of the blackbody plus power-law model versus mean count
rate (3.6--15.9 keV), with different observations of LMC X-1 indicated by
separate symbols.  The fitting was done over the 3.6--15.9 keV energy range,
on subsets of data selected by PCA count rate.}
\end{figure}

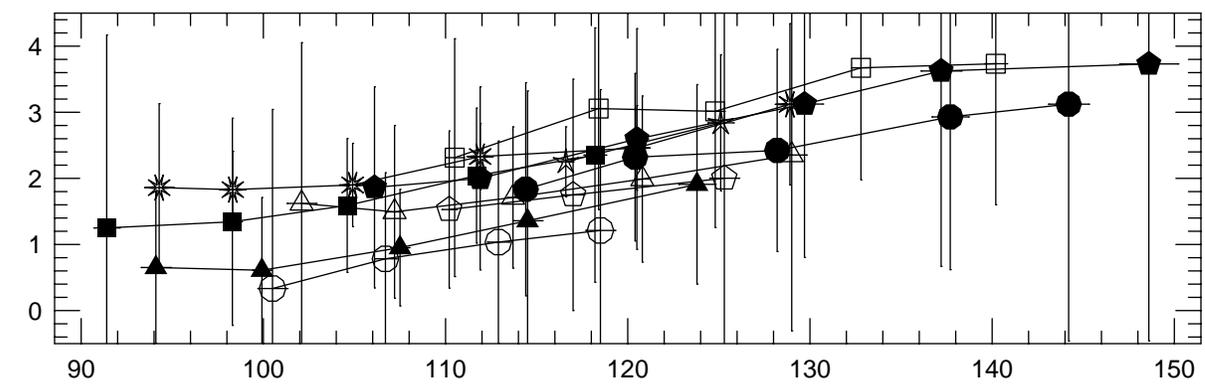
\end{document}